\documentclass{amsart}
\usepackage{amssymb}
\usepackage{amsfonts}
\usepackage{amsmath}
\usepackage{graphicx}%
\usepackage{lastpage}
\usepackage[legalpaper,bookmarks=true,colorlinks=true,linkcolor=blue,citecolor=blue]{hyperref}
\usepackage{fancyhdr}
\usepackage{color}
\usepackage[mathlines]{lineno}
\usepackage{lscape}
\usepackage{epsfig}
\usepackage{natbib}
\usepackage{float}

\newtheorem{theorem}{Theorem}
\theoremstyle{plain}

\numberwithin{equation}{section}

\begin{document}
\Large
\title[IF's of Theil-like Measure]{On the Influence Function for the Theil-like class of Inequality Measures}

\begin{abstract}
On one hand, a large class of inequality measures, which includes the generalized entropy, the Atkinson, the Gini, etc., for example,  has been introduced in Mergane and Lo (2013). On the other hand, the influence function of statistics is an important tool in the asymptotics of a nonparametric statistic. This function has been and is being determined and analysed in various aspects for a large number of statistics. We proceed to a unifying study of the \textit{IF} of all the members of the so-called Theil-like family and regroup those \textit{IF}'s in one formula. Comparative studies become easier.\\

\bigskip\noindent 
\author{$^{\dag}$ Tchilabalo Abozou Kpanzou}
\author{$^{\dag\dag}$ Diam Ba}
\author{$^{\dag\dag\dag}$ Pape Djiby Mergane}
\author{$^{\dag\dag\dag\dag}$ Gane Samb Lo}

\noindent $^{\dag}$ Tchilabalo Abozou Kpanzou (corresponding author).\\
Kara University, Kara, Togo\\
Email : kpanzout@gmail.com\\ 

\noindent $^{\dag\dag\dag}$ Diam Ba\\
LERSTAD, Gaston Berger University, Saint-Louis, S\'en\'egal.\\
Email : diamba79@gmail.com.\\

\noindent $^{\dag\dag}$ Pape Djiby Mergane\\
LERSTAD, Gaston Berger University, Saint-Louis, S\'en\'egal.\\
Email : mergane@gmail.com.\\

\noindent $^{\dag\dag\dag \dag}$ Gane Samb Lo.\\
LERSTAD, Gaston Berger University, Saint-Louis, S\'en\'egal (main affiliation).\newline
LSTA, Pierre and Marie Curie University, Paris VI, France.\newline
AUST - African University of Sciences and Technology, Abuja, Nigeria\\
gane-samb.lo@edu.ugb.sn, gslo@aust.edu.ng, ganesamblo@ganesamblo.net\\
Permanent address : 1178 Evanston Dr NW T3P 0J9,Calgary, Alberta, Canada.\\

\noindent \textbf{keywords and phrases}. Influence function, Measures of inequality, Lorenz curve, quantile function, Pareto law, Exponential law, Singh-Maddala law, lognormal law, .\\

\noindent \textbf{AMS 2010 Mathematics Subject Classification :} 62G35, 97K70.
\end{abstract}

\maketitle

\section{Introduction} 

\noindent Over the years, a number of measures of inequality have been developed. Examples include the generalized entropy, the Atkinson, the Gini, the quintile share ratio and the Zenga measures (see e.g. \cite{CowellandFlachaire2007}; \cite{Cowelletal2009}; \cite{HulligerandSchoch2009}; \cite{Zenga1984} and \cite{Zenga1990}). Recently, \cite{merganelo2013} gathered a significant number of inequality measures under the name of Theil-like family. Such inequality measures are very important in capturing inequality in income distributions. They also have applications in many other branches of Science, e.g. in ecology (see e.g. \cite{Magurran1991}), Sociology (see e.g. \cite{Allison1978}), Demography (see e.g. \cite{White1986}) and information science (see e.g. \cite{Rousseau1993}).\\

\noindent In order to make the above mentioned measures applicable, one often makes use of estimation. Classical methods unfortunately rely heavily on assumptions which are not always met in practice. For example, when there are outliers in the data, classical methods often have very poor performance. The idea in robust Statistics is to develop estimators that are not unduly affected by small departures from model assumptions, and so, in order to measure the sensitivity of estimators to outliers, the influence function (IF) was introduced (see \cite{Hampel1974}, \cite{Hampeletal1986}).\\

\noindent Let us begin by precising the objects and notation of our study, in particular the influence function. To make the reading of what follows easier, we suppose that we have a probability space $(\Omega, \mathcal{A}, \mathbb{E})$ holding a random variable $X$ associated with the cumulative distribution function (\textit{cdf}) $F(x)=\mathbb{P}(X\leq x)$, $x \in \mathbb{R}$, and a sequence of independent copies of $X$: $X_1$, $X_2$, etc. This random variable is considered as an income variable so that it is non-negative and $F(0)=0$. The absolute density distribution function (with respect to the Lebesgue measure on $\mathbb{R}$) of $X$ (\textit{pdf}), if it exists, is denoted by $f$. Its mean, we suppose finite and non-zero, and moments of order $\alpha\geq 1$ are denoted by 
$$
\mu_F =\int_{0}^{+\infty} y\ dF(y) \in (0,\infty) \ \ \text{and} \ \mu_{F,\alpha} =\int_{0}^{+\infty} y^{\alpha}\ dF(y), \ \mu_{F,1}=\mu_{F}.
$$ 
\bigskip \noindent The quantile function associated to $F$, also called generalized inverse function is defined by

$$
Q(p)\equiv F^{-1}(z)=\inf \{z \in \mathbb{R}, \ F(z)\leq x\}, \ p \in [0,1]
$$

\noindent and the Lorentz curve of $F$ is given by

$$
L(F,p)=\frac{q(p)}{\mu_F}, \ \text{with} \  q(p)=\int_{0}^{p} Q(s) \ ds, \  \ 0\leq s \leq 1.
$$

\noindent \noindent A nonparametric estimation $T(F)$ will studied as well as its plug-in nonparametric estimator of the form $T(F_n)$ which is based on the sample $X_1$,..., $X_n$, $n\geq 1$.\\

\noindent The influence function $IF(\circ,T(F))$ of $T(F)$ is the Gateaux derivative of $T$ at $F$  in the direction of Dirac measures in the form

\begin{equation}
IF(z,T(F))=\lim_{\epsilon\to 0}\frac{T(F_{\epsilon}^{(z)})-T(F)}{\epsilon}=\frac{\partial }{\partial\epsilon}T(F_{\epsilon}^{(z)})|_{\epsilon=0}, \ \ \label{IFformula}
\end{equation}

\bigskip \noindent where 

$$
F_{\epsilon}^{(z)}(u)=(1-\epsilon)F(u)+\epsilon\Delta_{Z}(u),\epsilon\in[0;1],
$$

\bigskip \noindent $\Delta_z$ is the \textit{cdf} of the $\delta_{z}$, the Dirac measure with mass one at $z$ and $z$ is in the value domain of $F$.\\

\noindent It is known that the asymptotic variance of the plug-in estimator $T(F_n)$ of statistic $T(F)$ is of the form $\sigma^2=\int IF(x,T(F))^2 dF(x)$ under specific condition, among them the Hadamard differentiability (see \cite{wasserman2006}, Theorem 2.27, page 19). So the influence function gives an idea of what might be the variance of the Gaussian limit of the estimator if it exists. At the same time, the behavior of its tails (lower and upper) give indications on how lower extreme and/or upper extreme values impact on the quality of the estimation. For example, recently, the sensitivity of a statistic $T(F)$ and the impact of extreme observations of some influence functions have been studied by, e.g., \cite{CowellandFlachaire2007}.\\

\noindent Another interesting fact is that the influence function behaves in nonparametric estimation as the score function does in the parametric setting (see \cite{wasserman2006}, page 19).\\

\noindent An area of application of the influence function is that of measures of inequality (see, e.g., \cite{VanPraagetal1983}, \cite{VictoriaFeser2000} and \cite{Kpanzou2015}). Due to the importance of that key element in nonparametric estimation in Econometric and welfare studies, a collection of inequality measures is being actively made. To cite a few, the \textit{IF}'s of the following measures are given in the Appendix section: the generalized entropy class of measures of inequality GE($\alpha$), where $\alpha>0$, the mean logarithmic deviation (MDL), the Theil Measure, the Atkinson Class of Inequality Measures of parameter $\alpha \in (0,1]$, the Gini Coefficient, the Quintile Share Ratio Measure of Inequality (QSR).\\

\noindent Fortunately, \cite{merganelo2013} introduced the so-called Theil-like family, in which are gathered the Generalized Entropy Measure, the Mean Logarithmic Deviation (\cite{cfa0203}, \cite{theil}, \cite{cfa80a}), the different inequality measures of Atkinson (\cite{atkinson}), Champernowne (\cite{champernowne}) and Kolm (\cite{kolm76}) in the following form:

\begin{equation}\label{ineq1}
T(F) =  \tau \left(\frac{1}{h_1\left(\mu_n\right)}\, \frac{1}{n} \sum_{j=1}^{n}h\left(X_j\right)\,-\,h_2\left(\mu_n\right)\right),\\
\end{equation}

\noindent where $\mu_n=\frac{1}{n}\sum_{j=1}^{n}X_j$ denotes the empirical mean while  $h$, $h_1$, $h_2,$ and $ \tau$ are measurable functions.\\

\bigskip \noindent The inequality measures mentioned above are derived from (\ref{ineq1}) with the particular values of $\alpha,  \tau, h, h_1$ and $h_2$ as described below for all $s > 0$ :

\begin{itemize}
	\item[(a)] Generalized Entropy
	$$ \alpha\neq 0,\, \alpha \neq 1,\;  \tau(s) = \frac{s-1}{\alpha\left(\alpha-1\right)},\,h(s)=h_1(s)=s^{\alpha},\; h_2(s)\equiv 0;$$
	\item[(b)] Theil's measure
	$$ \tau(s) = s,\; h(s)= s\,\log(s),\; h_1(s)=s,\; h_2(s)= \log(s);$$
	\item[(c)] Mean Logarithmic Deviation
	$$ \tau(s) = s,\; h(s)= h_2(s)=\log(s^{-1}),\; h_1(s)\equiv 1;$$
	\item[(d)] Atkinson's measure
	$$\alpha < 1 \textrm{ and } \alpha\neq 0,\;  \tau(s)=1-s^{1/\alpha},\; h(s)=h_1(s)=s^{\alpha},\; h_2(s)\equiv 0;$$
	\item[(e)] Champernowne's measure
	$$  \tau(s) = 1 - \exp\left(s\right),\; h(s)=h_2(s)=\log(s),\; h_1(s)\equiv 1;$$
	\item[(f)] Kolm's measure
	$$ \alpha > 0,\;  \tau(s) = \frac{1}{\alpha}\log(s),\; h(s)= h_1(s)=\exp (- \alpha s),\; h_2(s)\equiv 0.$$
\end{itemize}

\bigskip

\noindent This is simply the plug-in estimator of  

\begin{equation}
T(F)= \tau\left(\frac{\mathbb{E}h(X)}{h_1\left(\mu_F\right)} - h_2\left(\mu_F\right)\right)=\tau(I).
\end{equation}

\noindent The following conditions are required for the asymptotic theory.\\

\noindent \textbf{B1} The functions $\tau$ admits a derivative $\tau^{\prime}$ which is continuous at $I$ and $\tau^{\prime}(I)\neq 0$.

\noindent \textbf{B2}. The functions $h_1$ and $h_2$ admit derivatives $h_1^{\prime}$ and $h_2^{\prime}$ which are continuous at $\mu_F$ with $h_1(\mu_F)\neq 0$.\\

\noindent  \textbf{B3}. $\mathbb{E}h^j(X)<+\infty$, $j=1,2$.\\

\noindent This offers an opportunity to present a significant number of \textit{IF}'s in a unified approach. This may be an asset for inequality measures comparison. By the way, it constitutes the main goal of this paper.\\

\noindent Let us add more notation. The lower endpoint and upper endpoint of \textit{cdf} $F$ are denoted by
$$
lep(F)=\inf\{ y \in \mathbb{F}, \ F(x)>0\} \ \text{and} \ uep(F)=\sup\{y \in \mathbb{F}, \ F(x)<1\}.
$$ 

\noindent So the domain of admissible values for $X$, denoted by $\mathcal{V}_X$, satisfies $\mathcal{V}_X \subset \mathcal{R}_X=[lep(F), \ uep(F)]$, the latter being the range of $F$.\\

\noindent  The layout of this paper is as follows. In the next section we state our main result on the influence function of the TLIM family members and some particularized forms related to each known members. For member whose \textit{IF}'s are already given, we will make a comparison. In Section \ref{2018_07_11_sec_04}, we give the complete proofs. In Section \ref{2018_07_11_sec_05} we provide a conclusion and some perspectives. Section \ref{2018_07_11_sec_06} is an appendix gathering \textit{IF}'s expressions of some members of the TLIM available in the literature.

\section{Main results}

\noindent \textbf{(A) - The main theorem}.\\

\begin{theorem} \label{theo_IfTheilLike} If conditions $(B1)-(B2)$ hold, then the Influence function of the \textit{TLIM} index is given by

\begin{equation}
IF(z,F)=\tau^{\prime}(I) \biggr(- \left(\frac{h_1^{\prime}(\mu_F)\mathbb{E} h(X)}{h_1(\mu_F)^2}+h_2^{\prime}(\mu_F)\right) (z-\mu_F)  +\frac{h(X)-\mathbb{E}h(X)}{h_1(\mu_F)}\biggr),
\end{equation}

\noindent for $\ z \in \mathcal{V}_X$.
\end{theorem}

\noindent \textbf{Remark on the asymptotic variance}. It was said earlier that the plug-in estimator should give the asymptotic variance of the limiting Gaussian variable, if it exists, as 

$$
\sigma^2=\int_{\mathcal{V}_X} IF(X)^2 \ d\mathbb{P}=\mathbb{E}IF(X)^2.
$$

\bigskip \noindent This is exactly the case from the asymptotic normality of the plug-in estimator as established in Theorem 2 in \cite{merganeEtal2018}.\\

\noindent Let us move to the illustrations of our results for particular cases.

\bigskip \noindent \textbf{(B) - Particular forms}.\\

\noindent Let us proceed to the study of particular members of the TLIM class. We will have to compare our results with existent ones if any in the appendix. When the computation are simple, we only give the result without further details.\\

\noindent \textbf{(1) Mean Logarithmic Deviation}. We have 
$$
\tau(s) = s,\; h(s)= h_2(s)=\log(s^{-1}),\; h_1(s)\equiv 1
$$
\noindent and next $\tau^{\prime}(s) \equiv 1$, $h(s)^{\prime}= h_2^{\prime}(s)=-1/s$ and $h_1^{\prime}(s)\equiv 0$. The application of Theorem \ref{theo_IfTheilLike} leads to
$$
IF(z,DLM)=\mu_F^{-1} (z-\mu_F) + (\log z -\mathbb{E}\log X), \ z\in\mathcal{R}_F.
$$

\noindent \textbf{(2) Theil's Index}. We have
$$
\tau(s) = s,\; h(s)= s\log s, h_1(s)=s, \ \ h_2(s)=\log s,\; 
$$
\noindent and next $\tau^{\prime}(s) \equiv 1$, $h_1^{\prime}(s)\equiv 1$ and $h_2^{\prime}(s)=1/s$. The application of Theorem \ref{theo_IfTheilLike} gives
$$
IF(z,DLM)=\mu_F^{-1} (z\log z -\mathbb{E}X\log X) - \mu_F^{-2}(\mu_F +\mathbb{E}\log X), \ z\in\mathcal{R}_F.
$$

\noindent \textbf{(3) Class of Generalized Entropy Measures of parameter $\alpha$, $\alpha \notin \{0,1\}$}. We have
$$
\tau(s) = \frac{s-1}{\alpha\left(\alpha-1\right)}, \ \tau^{\prime}(s) = \frac{1}{\alpha\left(\alpha-1\right)},\, h(s)=h_1(s)=s^{\alpha}, h_1^{\prime}=\alpha s^{\alpha-1}\; h_2(s)\equiv 0.
$$
\noindent The application of Theorem \ref{theo_IfTheilLike} gives
$$
IF(z,GE(\alpha))=\frac{z^{\alpha}-\mu_{F,\alpha}}{\alpha(\alpha-1)\mu_{F}^{\alpha}}-{\mu_{F,\alpha}}{(\alpha-1)\mu_{F}^{\alpha+1}}(z-\mu_F), \ z\in\mathcal{R}_F.
$$ 

\noindent \textbf{(4) Class of Atkinson measures with parameter $\beta \in (0,1)$}. We have
$$
\tau(s)=1-s^{1/\beta},\; h(s)=h_1(s)=s^{\beta},\; h_2(s)\equiv 0.
$$
\noindent If we denote $\|X\|_{\beta}=\left(\mathbb{E}|X|^{\beta}\right)^{1/\beta}$, the application of Theorem \ref{theo_IfTheilLike} yields
$$
IF(z,At(\beta))=\frac{\|X\|_{\beta}}{\mu_{F}}\left(\frac{z-\mu_{F}}{\mu_{F}}-\frac{z^{\alpha}-\mu_{F,\beta}}{\beta \mu_{F}\mu_{F,\beta}}\right), \ z \in \mathcal{R}_F.
$$

\noindent \textbf{(5) Champernowne's index}. We have
$$
\tau(s) = 1 - \exp\left(s\right),\; h(s)=h_2(s)=\log(s),\; h_1(s)\equiv 1.
$$
\noindent The application of Theorem \ref{theo_IfTheilLike} implies that
$$
IF(z,Champ)=\frac{\exp(\mathbb{E}\log X)}{\mu_F} \left(\frac{1}{\mu_F}\left(z-\mu_F\right) - \left(\log z -\mathbb{E}\log X\right)\right), \ \ z \in \mathcal{R}_F.
$$

\noindent \textbf{(6) Kolm's Familily of inequality measure of parameter $\alpha \neq 0$}. We have
$$
\tau(s) = \frac{1}{\alpha}\log(s),\; h(s)= h_1(s)=\exp (- \alpha s),\; h_2(s)\equiv 0.
$$
\noindent By Theorem \ref{theo_IfTheilLike}, we have
$$
IF(z, \ Kolm(\alpha))=\frac{1}{\alpha_F} \left(\left(z-\mu_F\right) - \left(\frac{\exp(-\alpha \mu_F)}{\mathbb{E}\exp(-\alpha X)}-1\right)\right), \ z \in \mathcal{R}_F.
$$

$\diamond$\\

\section{Proof of the main theorem} \label{2018_07_11_sec_04}

\noindent In the following proof, we will use the method of finding the \textit{IF} following argument as given in \cite{kahn2015}. Suppose that we are interested in estimating $T(\mathbb{P}_X)$, where 
$\mathbb{P}_X$ the image measure is $d\mathbb{P}$ defined by $d\mathbb{P}_X(B)=d\mathbb{P}(X \in B)$ for $B\in \mathcal{B}(\mathbb{R})$ and is also Lebesgue-Stieljes probability law associated $F$, that is $\mathbb{P}_X(]a,b])=F(b)-F(a)$ for all $-\infty \leq a\leq b \leq +\infty$. Here we use integrals based on measures and thus integrals in $dF$ are integrals in $d\mathbb{P}_X$ in the following sense: for any non-negative and measurable function  $\ell : \mathbb{R} \rightarrow \mathbb{R}$, we have

$$
\int \ell(X) d\mathbb{P}=\inf h(y) d\mathbb{P}_X \equiv \int h(y) dF(y).
$$

\noindent Suppose that $T(\mathbb{P})$ is defined on a family of probability measures $\mathbb{P}_{\lambda}$, $\mathbb{P}_{\lambda}$ being associated with the random variable $X_{\lambda}$ with 
$X=X_{\lambda_0}$ and $F=F_{\lambda_0}$. Suppose that $T$ is independent of $\lambda$.  If we have

$$
\frac{\partial}{\partial \lambda} T(\mathbb{P}_{\lambda}) = \int \ell(y) \frac{\partial}{\partial \lambda} \mathbb{P}_{\lambda},
$$

\noindent where $\ell$ is measurable and $\mathbb{P}_X$-integrable. Then the IF at $T(F_{\lambda_0})=T(F)$ is given by

$$
IF(z,F)=\ell(z)- \int \ell(y) \ dF(y)=\ell(z)-\mathbb{E}\ell(X).
$$

\noindent Actually, the rule uses G\^ateaux differentiations properties and constitutes one of the fastest methods of finding the \textit{IF}. We are going to apply it.\\

\noindent \textbf{Proof of Theorem \ref{theo_IfTheilLike}}.\\

\noindent We remind the notation.

\begin{eqnarray*}
I=\frac{\mathbb{E}h(X)}{h_1(\mu_F)} - h_2(\mu_F).
\end{eqnarray*}

\noindent We have

\begin{eqnarray*}
\frac{\partial}{\partial \lambda} TLIM(\mathbb{P}_X)&=& \frac{\partial}{\partial \lambda}\tau\biggr(\biggr(\frac{1}{h_1\left(\int X d\mathbb{P}\right)} \biggr) \int h(X) d\mathbb{P} - h_1\left(\int X d\mathbb{P}\right)\biggr).
\end{eqnarray*}

\bigskip \noindent We get

\begin{eqnarray*}
\frac{1}{\tau^{\prime}(I)}TLIM(\mathbb{P}_X)&=& -\frac{h_1^{\prime}(\mu_F)\mathbb{E} h(X)}{h_1(\mu_F)^2} \int X \frac{\partial}{\partial \lambda} d\mathbb{P}\\
&+& \frac{1}{h_1(\mu_F)} \int h(X) \frac{\partial}{\partial \lambda} \mathbb{P}\\
&-& h_2^{\prime}(\mu_F) \int X \frac{\partial}{\partial \lambda} d\mathbb{P}\\
&=&\int \biggr(-\left(\frac{h_1^{\prime}(\mu_F)\mathbb{E} h(X)}{h_1(\mu_F)^2}+h_2^{\prime}(\mu_F)\right) X +\frac{h(X)}{h_1(\mu_F)}\biggr) \frac{\partial}{\partial \lambda} d\mathbb{P}.
\end{eqnarray*}

\noindent By centering at expectations, we have

$$
IF(z,F)=\tau^{\prime}(I) \biggr(-\left(\frac{h_1^{\prime}(\mu_F)\mathbb{E} h(X)}{h_1(\mu_F)^2}+h_2^{\prime}(\mu_F)\right) (z-\mu_F) +\frac{h(X)-\mathbb{E}h(X)}{h_1(\mu_F)}\biggr), \ z \in \mathcal{V}_X.
$$

$\square$\\

\section{Conclusion and Perspectives} \label{2018_07_11_sec_05}
I this paper, we studied the Theil-like family of inequality measures introduced in \cite{merganeEtal2018}. Following the paper on the asymptotic finite-distribution normality, we focus on the influence function of that family. Results are compared with those of some authors in particular. We think that this unified and compact approach will serve as general tools for comparison purpose. In addition, in computation packages, it allows more compact programs resulting in more efficiency. A paper on computational aspects will follow soon.

\newpage
\section{Appendix: A list of some Influence Functions}\label{2018_07_11_sec_06}

\noindent Here, we list a number of inequality measures and the corresponding \textit{influence functions}.\\
 
\noindent \textbf{The Generalized Entropy Measures of Inequality GE($\alpha$)}, which depends of a parameter $\alpha>0$ and defined by

\begin{eqnarray*}\label{genentrmeooineq}
I_E^\alpha&=&\int_0^\infty\frac{1}{\alpha(\alpha-1)}\left[\left(\frac{y}{\mu_F}\right)^\alpha-1\right]dF(y)\\
&=& \frac{1}{\alpha(\alpha-1)}\left(\frac{\mu_{F,\alpha}}{\mu_F^\alpha}-1\right), \ \alpha>0, \ \alpha \notin \{0,1\},
\end{eqnarray*}

\bigskip \noindent has the \textit{IF} (see e.g. \cite{CowellandFlachaire2007})

\begin{equation}
IF(z;I_E^\alpha)=\textcolor{blue}{\frac{1}{\alpha(\alpha-1)\mu_F^\alpha}}(z^\alpha-\mu_\alpha)-\frac{\mu_\alpha}{(\alpha-1)\mu_F^{\alpha+1}}[z-\mu_F],\alpha\notin \{0,1\}.
\end{equation}

\noindent \textbf{Important remark}. Our result on the \textit{IF} of the $GE(\alpha)$ is different from that of \cite{CowellandFlachaire2007} by the multiplicative coefficient $\textcolor{blue}{\frac{1}{\alpha(\alpha-1)\mu_F^\alpha}}$. In other words, that coefficient is missing in \cite{CowellandFlachaire2007}. We also find the same result by the computations below which is a direct proof.

\begin{eqnarray*}
\frac{\partial }{\partial \lambda} GE(\alpha)&=& \frac{\partial }{\partial \lambda} GE(\alpha)= \frac{\partial }{\partial \lambda}\frac{1}{\alpha(\alpha-1)} \left(\frac{\int X^{\alpha} \ d\mathbb{P}}{\left(\int X \ d\mathbb{P}\right)^{\alpha}}-1\right)\\
&=& \frac{1}{\alpha(\alpha-1)} \int \frac{\mu_F^{\alpha}X^{\alpha}-\alpha \mu_F^{\alpha-1}X}{\mu_F^{2\alpha}} \frac{\partial }{\partial \lambda} d\mathbb{P}.\\
\end{eqnarray*}

\noindent By the method described in the proof, we may center the integrand to get 

$$
IF(X,GE(\alpha)=\frac{1}{\alpha(\alpha-1)} \frac{\mu_F^{\alpha}(X^{\alpha}-\mathbb{E}X^{\alpha})-\alpha \mu_F^{\alpha-1}(X-\mathbb{E}X)}{\mu_F^{2\alpha}}.
$$

\noindent which again gives the result. 

\bigskip \noindent \textbf{The Mean Logarithmic Deviation (MDL)}, which is a special case of the GE class where $\alpha=0$, defined by 
\begin{equation}
I_E^0=-\int_0^\infty \log\left(\frac{y}{\mu_F}\right)dF(y)=\log \mu_1-\nu, \ \nu=\mathbb{E}\log X,
\end{equation}

\bigskip \noindent is associated to the \textit{IF}

\begin{equation}IF(z,I_E^0)=-[\log z-\nu]+\frac{1}{\mu_1}[z-\mu_F].
\end{equation}

\bigskip 
\noindent \textbf{The Theil Measure}, which also is a special case of the GE class for $\alpha=1$, 

\begin{equation}
I_E^1=\int_0^\infty \frac{y}{\mu_F}\log\left(\frac{y}{\mu_F}\right)dF(y)=\frac{\nu}{\mu_F}-\log\mu_F, \ \ \nu=\mathbb{E}X\log X,
\end{equation}

\bigskip \noindent has the \textit{IF}

\begin{equation}
IF(z;I_E^1)=\frac{1}{\mu_F}[z\log z-\nu]-\frac{\nu+\mu_F}{\mu_1^2}[z-\mu_F].
\end{equation}

\bigskip \noindent \textbf{The Atkinson Class of Inequality Measures} of parameter $\alpha \in (0,1]$, defined by (see \cite{CowellandFlachaire2007})

\begin{eqnarray*} 
I_A^\alpha&=&1-\left[\int_0^\infty \left(\frac{y}{\mu_F}\right)^{1-\alpha}dF(y)\right]^{1/(1-\alpha)}\\
&=&1-\frac{\mu_{F,1-\alpha}^{1/(1-\alpha)}}{\mu_F}, \ \alpha>0, \alpha\neq1,
\end{eqnarray*}

\bigskip \noindent and its influence function is given by
\begin{equation}
IF(z;I_A^\alpha)=-\frac{\nu^{(1/(1-\epsilon))-1}}{(1-\epsilon)\mu_F}(z^{1-\epsilon}-\nu)+\frac{\nu^{1/(1-\epsilon)}}{\mu_F^2}(z-\mu_F), 
\end{equation}

\noindent where $\nu=\mathbb{E}X^{1-\epsilon}$.\\

\bigskip \noindent We notice that for $\alpha=1$, we have

\begin{equation}\label{atkinson1ineqmeas} I_A^1=1-\frac{e^{\int_0^\infty(\log y)dy}}{\mu}=1-e^{-I_E^0},
\end{equation}

\bigskip 
\noindent \textbf{The Gini Coefficient}, defined by (see e.g. \cite{CowellandFlachaire2007}):
\begin{equation} \label{giniwithlorcrv} 
I_G=1-2\int_0^1 L(F,p)dp,
\end{equation}

\bigskip 
\noindent has the \textit{IF}
\begin{equation}
IF(z,I_G)=2\left[R(F)-C(F,F(z))+\frac{z}{\mu_F}(R(F)-(1-F(z)))\right],
\end{equation}

\noindent where
\begin{equation}
R(F)=\int_0^1 L(F,p) \ dp
\end{equation}

\noindent and $C$ is is the cumulative functional defined by

\begin{equation}\label{cumfunctnal} 
C(F,p)=\int_0^{Q(p)}xdF(x), \ 0\leq p \leq 1.
\end{equation}

\bigskip \noindent \textbf{The Quintile Share Ratio Measure of Inequality (QSR)}, defined by

\begin{equation}
\label{tchiladefinitionofqsr}\eta=\frac{\int_{Q(0.8)}^\infty ydF(y)}{\int_0^{Q(0.2)}ydF(y)}=\frac{EX\mathbf{1}_{\{X>Q(0.8)\}}}{EX\mathbf{1}_{\{X\leq Q(0.2)\}}},
\end{equation}

\noindent where $\mathbf{1}_A$ is an indicator function of a set $A$, is associated with the \textit{IF} described below (see \cite{Kpanzou2015}). Let 
\begin{equation}\label{qsrnumrtr} 
N(F)=\int_{Q(0.8)}^\infty xdF(x)
\end{equation}

\noindent and
\begin{equation}\label{qsrdenomtr}
D(F)=\int_0^{Q(0.2)}xdF(x).
\end{equation}

\bigskip \noindent and define the subdivision  of $\mathbb{R}_+$ : $A_1=[0, \ Q(0.2)]$, $A_2=(Q(0.2), \ Q(0.8))$, $A_3=(Q(0.8), \ 1]$ and set

\begin{eqnarray*}
I_1(z,\eta)&=&-zN(F)+0.2Q(0.8)D(F)+0.8Q(0.2)N(F)]/D^2(F);\\
I_2(z,\eta)&=&0.2 \ Q(0.8)D(F)-0.2Q(0.2)N(F)]/D^2(F);\\
I_3(z,\eta)&=&zD(F)-0.8Q(0.8)D(F)-0.2Q(0.2)N(F)]/D^2(F).\\
\end{eqnarray*}

\bigskip \noindent The \textit{SQR} influence function is defined by

$$
I_1(z,\eta)=I_1(z,\eta) \mathbf{1}_{A_1}(z) + I_2(z,\eta) \mathbf{1}_{A_2}(z) + I_3(z,\eta) \mathbf{1}_{A_3}(z).
$$

\end{document}